\documentclass[useAMS,usenatbib,usegraphicx]{mn2e}
\usepackage{float}
\newcommand{\bugref}{\bibitem[\protect\citeauthoryear{dummy }{1893}]{dum}}

\title[Faraday-Rotation Structure Across AGN Jets]
{Parsec Scale Faraday-Rotation Structure Across the Jets of 9 Active 
Galactic Nuclei}

\author[D. C. Gabuzda et al.]
{D. C. Gabuzda, N. Roche, A. Kirwan, S. Knuettel, M. Nagle \& C. Houston \\
Physics Department, University College Cork, Cork, Ireland} 

\begin{document}

\date{}
\pagerange{\pageref{firstpage}--\pageref{lastpage}} \pubyear{2013}
\maketitle
\label{firstpage}
\begin{abstract}
A number of groups have recently been active in searching for gradients in
the observed Faraday rotation measure (RM) across jets of Active Galactic Nuclei
(AGNs) on various scales and estimating their reliability. Such RM structures
provide direct evidence for the presence of an azimuthal magnetic field 
component, which may be associated with a helical jet magnetic field, as is
expected based on the results of many theoretical studies.  
We present new parsec-scale
RM maps of 4 AGNs here, and analyze their transverse RM structures together
with those for 5 previously published RM maps. All these jets display
transverse RM gradients with significances of at least $3\sigma$. This is 
part of an ongoing effort to establish how common transverse RM gradients 
that may be associated with helical or toroidal magnetic fields are in AGNs 
on parsec scales. 
\end{abstract}

\section{Introduction}

The relativistic jets of Active Galactic Nuclei (AGNs) emit radio synchrotron
emission, which can be linearly polarized up to about 75\% in optically thin 
regions with uniform magnetic fields, with the polarization angle $\chi$ 
orthogonal to the projection of the magnetic field {\bf B} onto the plane of 
the sky (Pacholczyk 1970). The degree of linear polarization is considerably
lower in optically thick regions,  up to 10--15\%, with $\chi$ parallel to
the projected {\bf B} (Pacholczyk 1970).  

In the standard theoretical picture of AGN jets offered by Blandford \& 
K\"onigl (1979), emission is observed only near and beyond the location 
in the jet outflow where the optical depth is near unity, $\tau \approx 1$, 
representing the transition between optically thick regions closer to the 
central engine and optically thin regions farther along the jet. This 
$\tau \approx 1$ surface is a theoretical construction that is located 
somewhere within the ``core'' observed in Very Long Baseline Interferometry 
(VLBI) images. Although there is a tendency to think broadly in terms of 
an optically thick core and optically thin jet, there is abundant evidence 
that the VLBI cores observed at centimeter wavelengths are in fact mixed 
regions of partially optically thick emission corresponding to the vicinity of
the theoretical $\tau \approx 1$ surface and optically thin emission from the 
inner jet (see, e.g., Gabuzda 2015).

\begin{center}
\begin{table*}
\begin{tabular}{c|c|c|c|c|c}
\multicolumn{6}{c}{Table 1: Source properties}\\
\hline
Source & Alternate & Redshift & pc/mas &  Integrated RM &  Original RM   \\ 
       & Name &         &        &  (rad\,m$^{-2}$)  &   Map Ref \\\hline
0212+735   & S5 0212+73 & 2.37   & 8.28   &    $+22\pm 1$  &   2      \\
0300+470   & 4C +47.08  &---    & ---    &    $+22\pm 6$  &   * \\
0305+039  & 3C 78 & 0.029 & 0.57 &   $+10\pm 2$  &   3 \\
0415+379   & 3C 111 & 0.049  & 0.95   &    $-24\pm 11$  &   1     \\ 
0945+408   & 4C +40.24   & 1.249  & 8.50   &    $+3\pm 1$   &    *\\ 
1502+106   & OR 103   & 1.838  & 8.53   &    $+1\pm 2$  &    *     \\
1611+343   & DA406   & 1.4    & 8.50   &    $+15\pm 1$  &    2     \\
2005+403   & TXS 2005+403 & 1.736  & 8.55   &    $-171\pm 12$  &    2     \\ 
2200+420   & BL~Lac & 0.069  & 1.29    & $-199\pm 1$  &    *   \\ \hline  
\multicolumn{6}{l}{1 = Zavala \& Taylor (2002); 2 = Zavala \& Taylor (2003); 
3 = Kharb et al. 2009;}\\
\multicolumn{6}{l}{* = This RM map not previously published}\\
\end{tabular}
\end{table*}
\end{center}

\begin{center}
\begin{table*}
\begin{tabular}{c|c|c|c|c|c|c|c|c|c}
\multicolumn{10}{c}{Table 2: Map properties}\\
\hline
Source & Epoch & Figure & Freq & Beam &  Peak  & Lowest contour & BMaj & BMin & BPA \\ 
      &    &  & (GHz)   & Shape & (Jy/beam) &  (\%) & (mas) & (mas) & (deg) \\\hline
0212+735 & June 27, 2000 & \ref{fig:rmmaps-ZT}   & 8.1 &  E & 2.32  & 0.50 & 1.50  & 1.03  &  $-7.4$  \\
0212+735 & June 27, 2000 & \ref{fig:rmmaps-ZT}   & 8.1 &  C   & 2.39 & 0.50 & 1.24  & 1.24  &   --     \\\hline
0300+470 & Sept. 26, 2007 & \ref{fig:rmmaps-BG173-1}   & 4.6 &  E   & 0.60 & 0.25 & 3.41  & 1.86  & $-7.9$\\
0300+470 & Sept. 26, 2007 & \ref{fig:rmmaps-BG173-1}   & 4.6 &  C & 0.60   & 0.25 & 2.50  & 2.50  &  --  \\\hline
0305+039 & Sept. 10, 2005 & \ref{fig:rmmaps-PK}   & 5.0  &  E  &  0.34 & 0.25 & 3.30  & 1.38  & $-8.1$   \\
0305+039 & Sept. 10, 2005 & \ref{fig:rmmaps-PK}   & 5.0  &  C   & 0.37 & 0.25 & 2.06  & 2.06  &  --  \\\hline
0415+379 & June 27, 2000 & \ref{fig:rmmaps-ZT}  & 8.1 &  E& 0.83    & 0.25 & 1.99  & 0.98  &  $-2.8$ \\
0415+379 & June 27, 2000 & \ref{fig:rmmaps-ZT}   & 8.1 &  C& 0.92    & 0.25 & 1.40  & 1.40  &  --     \\\hline
0945+408 & Sept. 26, 2007 & \ref{fig:rmmaps-BG173-1} & 4.6 &  E& 0.78   & 0.50 & 3.98  & 2.07  & $-9.5$  \\
0945+408 & Sept. 26, 2007 & \ref{fig:rmmaps-BG173-1} & 4.6 &  C& 0.85   & 0.50 & 2.87  & 2.87  & --      \\\hline
1502+106 & Sept. 26, 2007 & \ref{fig:rmmaps-BG173-1} & 4.6 &  E& 0.98   & 0.25 & 5.30  & 1.95  & $-18.0$\\
1502+106 & Sept. 26, 2007 & \ref{fig:rmmaps-BG173-1} & 4.6 &  C& 1.03   & 0.25 & 3.20  & 3.20  & --    \\\hline
1611+343 & June 27, 2000 & \ref{fig:rmmaps-ZT}   & 8.1 &  E& 2.70   & 0.50 & 2.00  & 0.97  & $-5.0$  \\
1611+343 & June 27, 2000 & \ref{fig:rmmaps-ZT}   & 8.1 &  C& 2.66   & 0.50 & 1.40  & 1.40  & --      \\\hline
2005+403 & June 27, 2000 & \ref{fig:rmmaps-ZT}   & 8.1 &  E  & 1.02 & 1.00 & 2.31  & 1.47  & $-14.0$  \\
2005+403 & June 27, 2000 & \ref{fig:rmmaps-ZT}   & 8.1 &  C   & 1.12& 1.00 & 1.84  & 1.84  &  --     \\\hline
2200+420 & Sept. 26, 2007 & \ref{fig:rmmaps-BG173-2}   & 4.6 &  E& 1.75   & 0.50 & 3.97  & 1.69  & $-23.7$ \\
2200+420 & Sept. 26, 2007 & \ref{fig:rmmaps-BG173-2}   & 4.6 &  C& 1.75   & 0.50 & 2.59  & 2.59  & --      \\
2200+420 & Sept. 26, 2007 & \ref{fig:rmmaps-BG173-2}   & 4.6 &  E& 1.54   & 1.00 & 2.27  & 1.15  & $-24.3$ \\
2200+420 & Sept. 26, 2007 & \ref{fig:rmmaps-BG173-2}   & 4.6 &  C& 1.56   & 1.00 & 1.60  & 1.60  & --      \\ \hline
\end{tabular}
\end{table*}
\end{center}

Multi-frequency VLBI polarization
observations provide  information about the wavelength dependence of
the parsec-scale polarization, in particular, Faraday rotation occurring 
at various locations between the emitting region and observer.  
When Faraday rotation occurs in regions of thermal (non-relativistic
or only mildly relativistic) plasma outside 
the emitting region the rotation is given by
\begin{eqnarray}
           \chi_{obs} - \chi_o = 
\frac{e^3\lambda^{2}}{8\pi^2\epsilon_om^2c^3}\int n_{e} 
{\mathbf B}\cdot d{\mathbf l} \equiv RM\lambda^{2}
\end{eqnarray}
where $\chi_{obs}$ and $\chi_o$ are the observed and intrinsic 
polarization angles, respectively, $-e$ and $m$ are the charge and 
mass of the particles giving rise to the Faraday rotation, usually 
taken to be electrons, $c$ is the speed of light, $\epsilon_o$ the
permittivity constant, $n_{e}$ the 
density of the Faraday-rotating electrons, $\mathbf{B}$ the magnetic 
field, $d\mathbf{l}$ an element along the line of sight, $\lambda$ 
the observing wavelength, and RM (the coefficient of $\lambda^2$) is the 
Rotation Measure (e.g., Burn 1966).  The action of external Faraday rotation 
can be identified using simultaneous multifrequency observations, through 
the linear $\lambda^2$ dependence, allowing the determination of both the 
RM (which reflects the electron density and line-of-sight {\bf B} field 
in the region of Faraday rotation) and $\chi_o$ (the intrinsic direction
of the source's linear polarization, and hence the synchrotron {\bf B} field, 
projected onto the plane of the sky).

Many theoretical studies and simulations of the relativistic jets of AGNs
have predicted the development of a helical jet {\bf B} field, which comes
about due to the combination of the rotation of the central black hole and
its accretion disk and the jet outflow (e.g. Nakamura, Uchida \& Hirose 2001, 
Lovelace et al. 2002; see Tchekhovskoy and Bromberg 2016 for a recent
example). Researchers have long been aware that the presence of a helical
jet {\bf B} field could give rise to a regular gradient in the observed 
RM across the jet, due to the systematic change in the line-of-sight 
component of the helical field (Perley et al. 1984, Blandford 1993). 
Statistically significant transverse RM gradients across 
the parsec-scale jets of more than 25 AGN have been reported in the
refereed literature (e.g. Gabuzda et al. 2015 and 
references therein), interpreted as reflecting the systematic change in the 
line-of-sight component of a toroidal or helical jet {\bf B} field across 
the jets.

The Monte Carlo simulations of Hovatta et al. (2012), Mahmud et al.
(2013) and Murphy \& Gabuzda (2013) clearly indicate that the key factors in
determining the trustworthiness of an RM gradient (i.e., the probability
that it is not spurious) are (i) monotonicity, (ii) the range of values 
encompassed by the gradient relative to the uncertainties in the RM 
measurements and (iii) steadiness of the change in the RM values across 
the jet (ensuring a seeming ``gradient'' is not due only to values in a 
few edge pixels), rather than the width spanned by the gradient. The second
criterion reflects the result of the Monte Carlo simulations that, when
an RM gradient encompasses values differing by at least $3\sigma$ and spans 
even a small distance comparable to one beamwidth, the probability
that it is spurious is very low -- less than 1\%.
 

Mahmud et al. (2013), Gabuzda et al. (2014a, 2014b) and Motter and
Gabuzda (2017) have carried out new Faraday-rotation analyses employing 
the empirical error formula of
Hovatta et al. (2012), focusing on monotonicity, steadiness of the 
gradient across the jet and a significance of
at least $3\sigma$ as the key criteria for reliability of observed
transverse RM gradients. Results published earlier by Gabuzda et al. 
(2004, 2008) were reanalyzed using this same approach by Gabuzda et 
al. (2015), who also reported 8 new cases of monotonic, statistically 
significant transverse RM gradients across AGN jets based on 
previously published and unpublished maps.

In the current study, we have applied this
approach 
to analyze 4 RM images 
previously published by Zavala \& Taylor (2002, 2003), based on VLBA data
at 7 frequencies between 8.1 and 15.2~GHz, 1 RM image published by Kharb
et al. (2009) based on 3 frequencies between 5.0 and 15.3~GHz  
and 4 RM images published here for the first time, based on 
VLBA data at 6 frequencies
between 4.6 and 15.4~GHz. All of these AGNs display statistically
significant transverse RM gradients across their jets.
This is part of a larger
study aiming to build up statistics for AGN jets displaying transverse
RM gradients with the ultimate goal of analyzing the collected 
properties of the RM gradients detected.

\begin{figure*}
\begin{center}
\includegraphics[width=0.33\textwidth,angle=90]{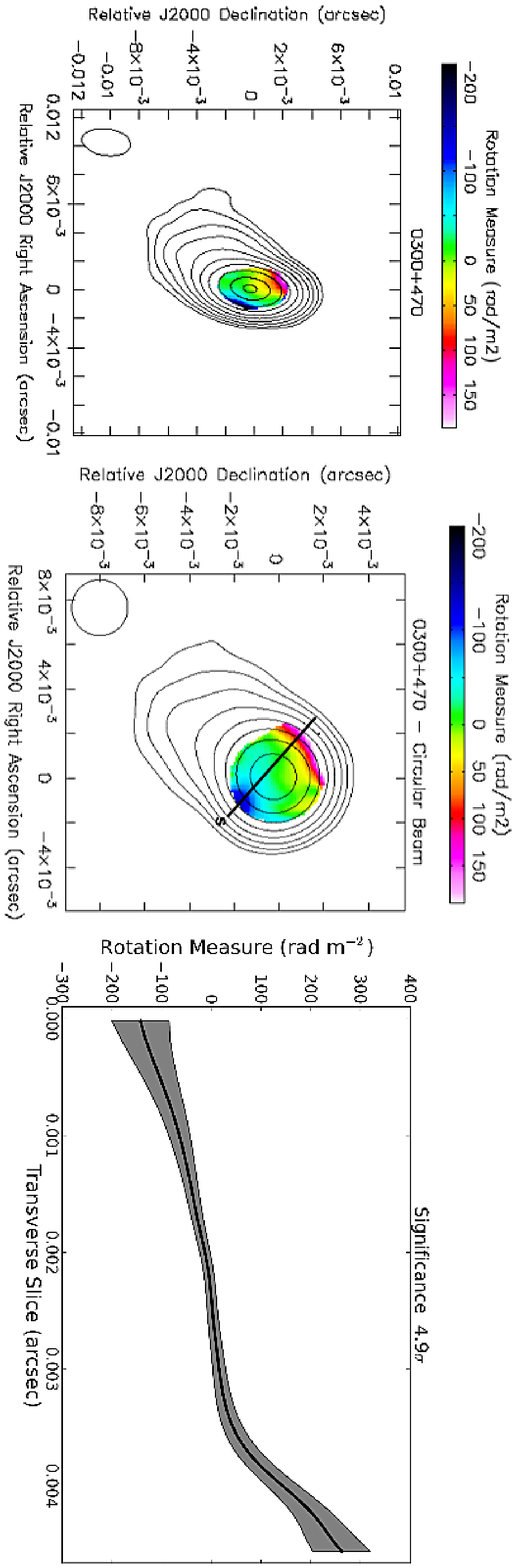}
\includegraphics[width=0.31\textwidth,angle=90]{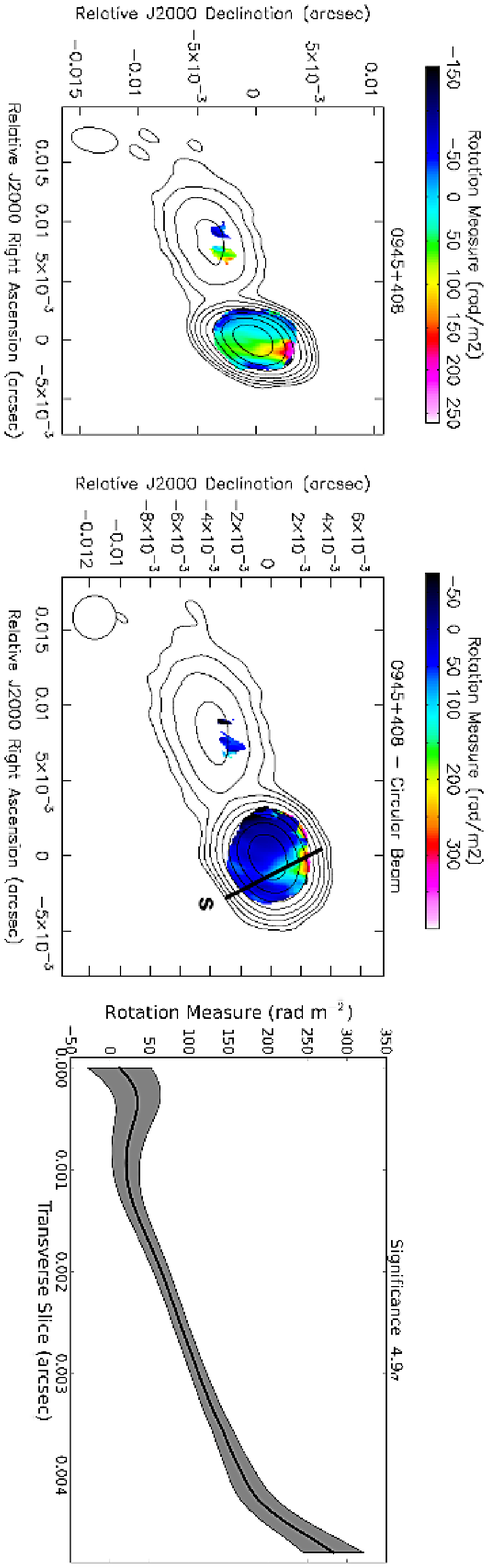}
\includegraphics[width=0.50\textwidth,angle=90]{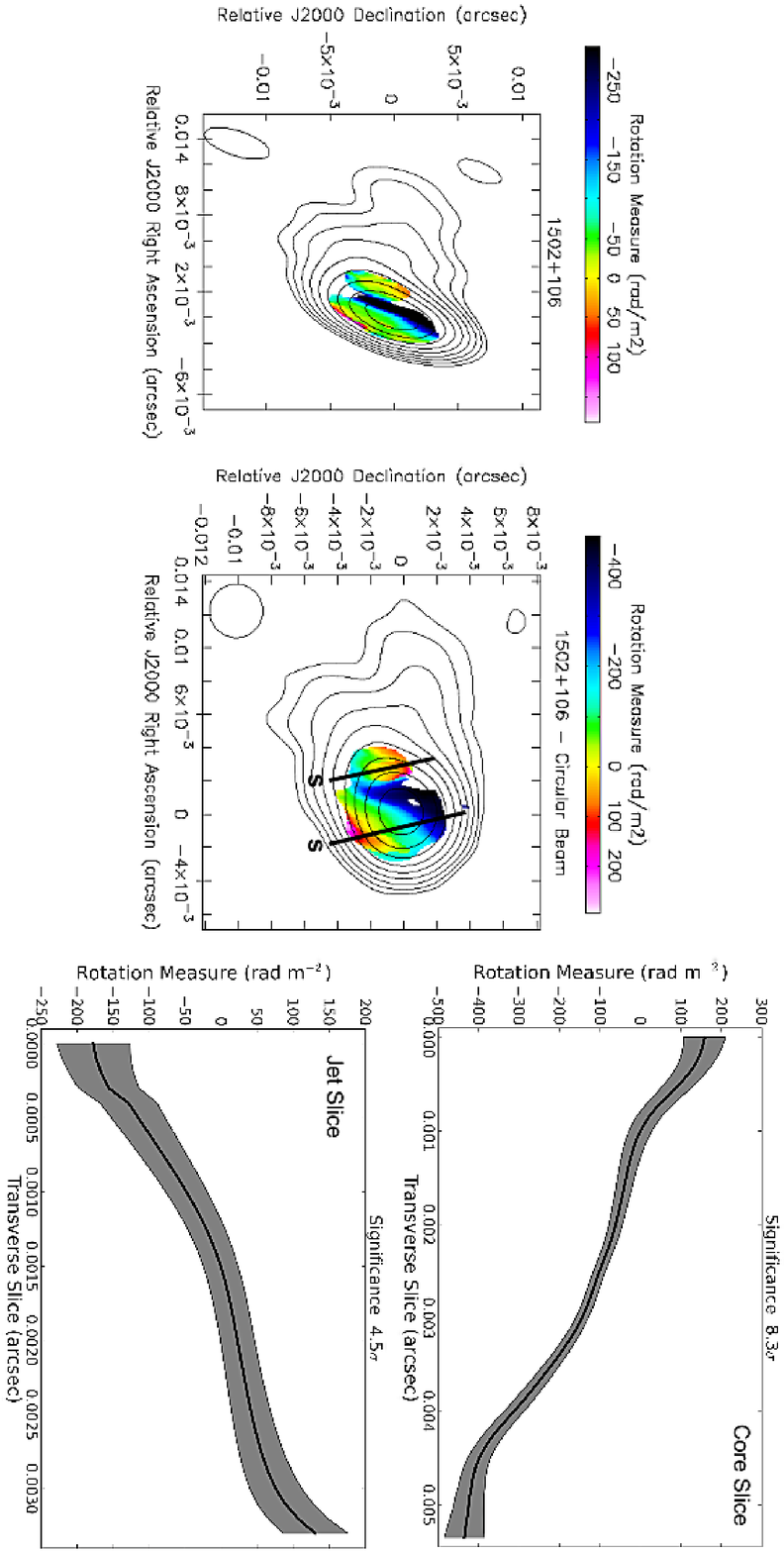}
\end{center}
\caption{4.6-GHz intensity maps and transverse Faraday RM gradients across 
the jets of 0300+470 (top), 0945+408 (middle) and 1502+106 (bottom), based 
on the 4.6--15.4~GHz data described in Section~2.1.  
}
\label{fig:rmmaps-BG173-1}
\end{figure*}

\begin{figure*}
\begin{center}
\includegraphics[width=0.75\textwidth,angle=90]{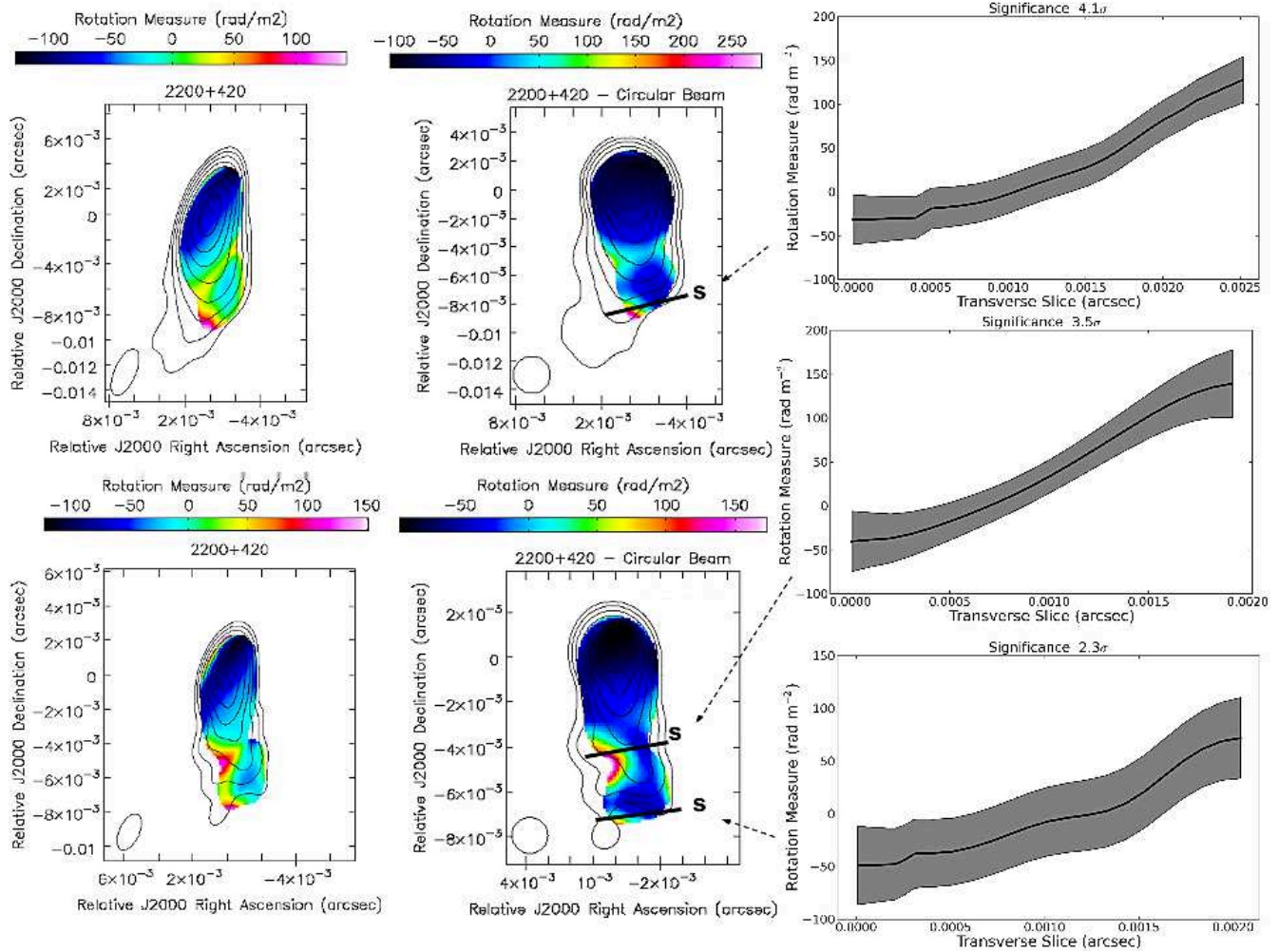}
\end{center}
\caption{Transverse Faraday RM gradient across the jet of 2200+420, based on
the 4.6--15.4~GHz data described in Section~2.1. 
The upper row of maps was made using the 4.6~GHz beam, 
while the lower row of maps correspond to the same data convolved with the
(slightly smaller) 7.9-GHz beam.  
}
\label{fig:rmmaps-BG173-2}
\end{figure*}

\begin{figure*}
\begin{center}
\includegraphics[width=0.31\textwidth,angle=90]{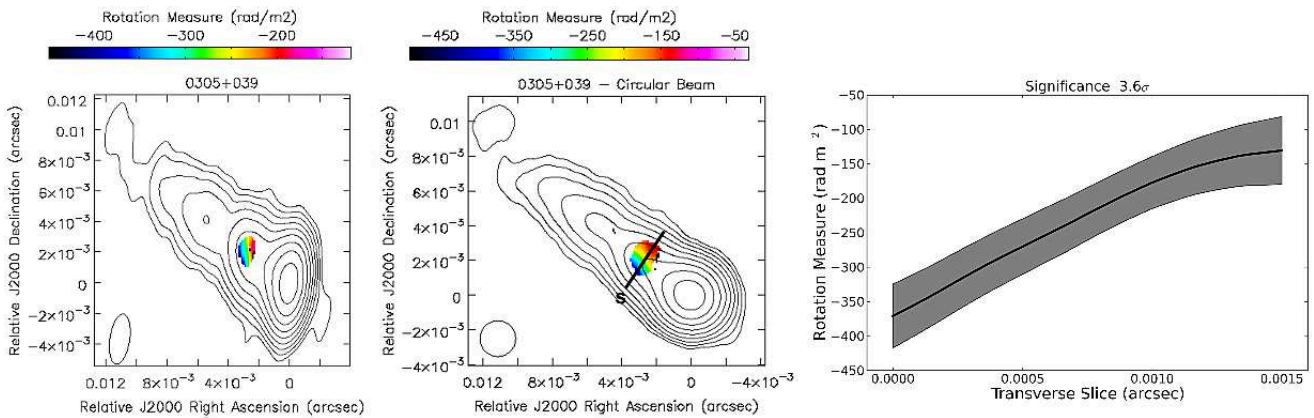}
\end{center}
\caption{5.0-GHz intensity maps and and transverse Faraday RM gradient 
across the jet of 0305+039, based on the 5.0--15.3~GHz data described 
in Section~2.2. 
}
\label{fig:rmmaps-PK}
\end{figure*}

\begin{figure*}
\begin{center}
\includegraphics[width=0.31\textwidth,angle=90]{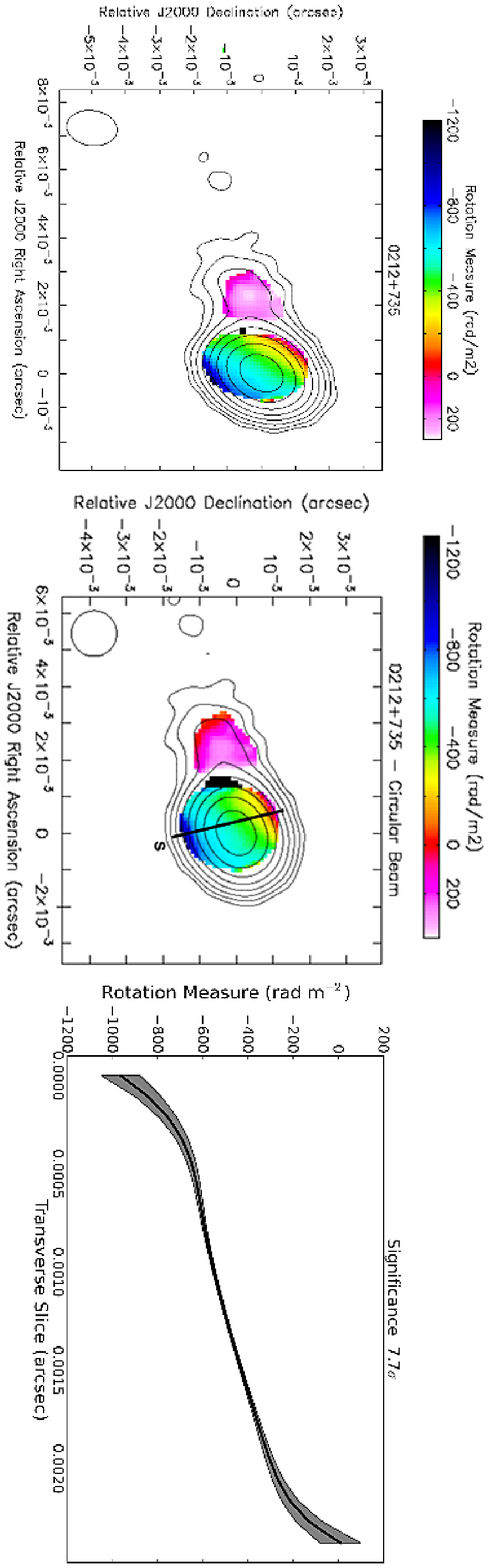}
\includegraphics[width=0.33\textwidth,angle=90]{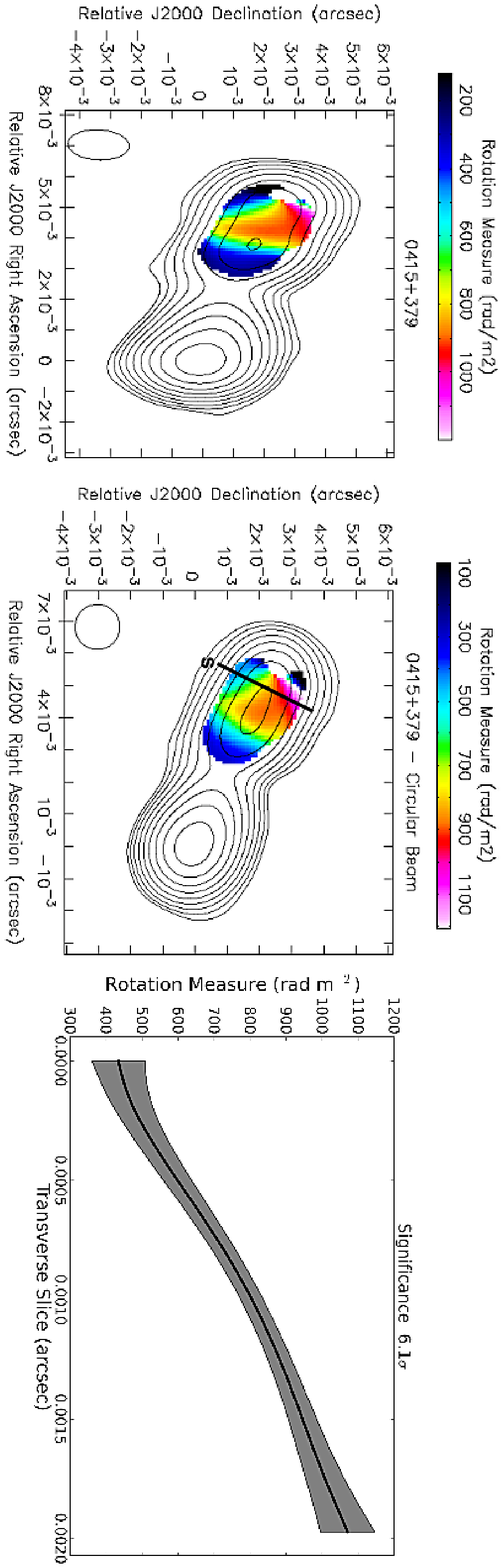} 
\includegraphics[width=0.33\textwidth,angle=90]{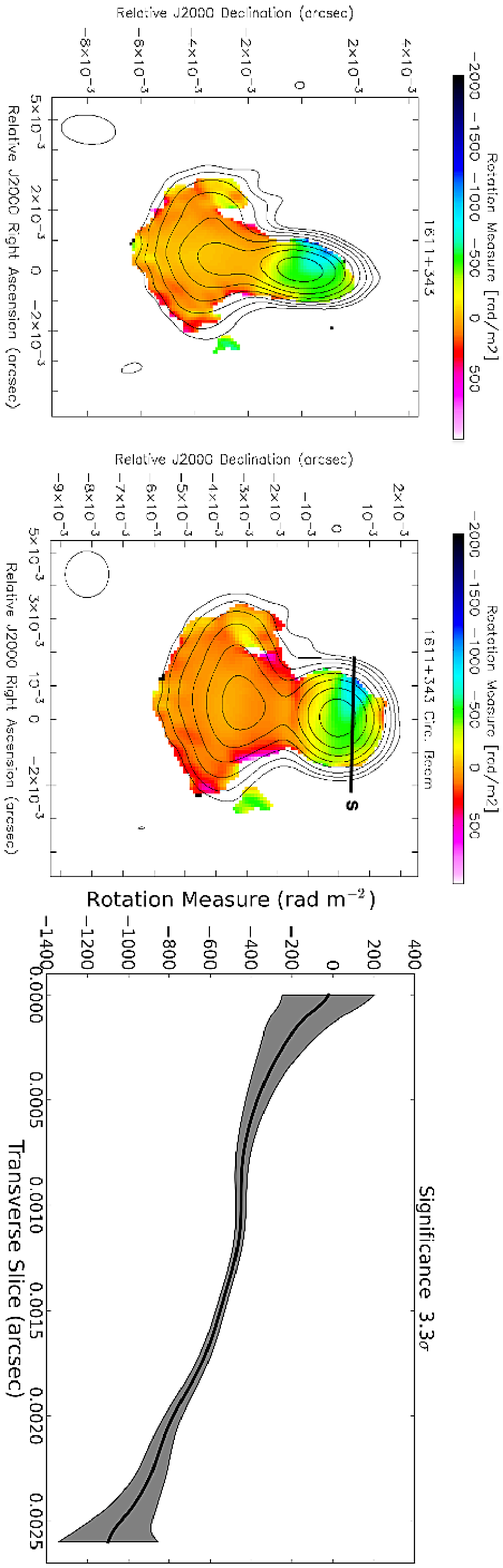}
\includegraphics[width=0.31\textwidth,angle=90]{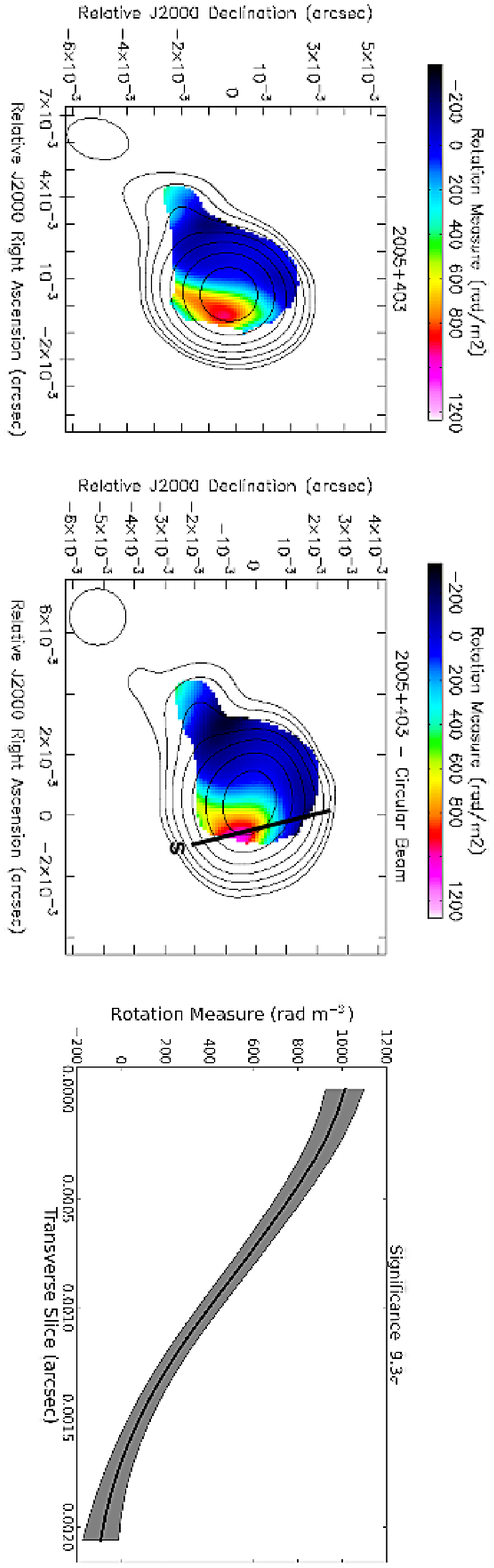} 
\end{center}
\caption{8.1~GHz intensity maps and transverse Faraday RM gradients across 
the jets of 0212+735, 0415+379, 1611+343 and 2005+403, from top to bottom,
based on the 8.1--15.2~GHz data described in Section~2.3. 
}
\label{fig:rmmaps-ZT}
\end{figure*}

\section{Observations }

Like Gabuzda et al. (2015), we present some new analyses of previously 
published Faraday RM maps together with RM images published here for the first 
time.  In all cases, 
the observations were obtained on the NRAO Very Long Baseline Array.
We carried out the imaging and analysis for the data considered here
in the same way as is described by Gabuzda et al. (2015), including
matching the resolutions at the different frequencies, aligning the images
at the different frequencies when significant misalignments were present,
and correction for Faraday rotation occurring in our Galaxy when significant.


The integrated RM measurements of Taylor et al. (2009) for all the sources 
considered here, based on the VLA Sky Survey (NVSS) observations at two
bands near 1.4~GHz, are given in Table~1.  The integrated RMs for 0212+735, 
0300+470, 0305+039, 0415+379, 0945+408, 1502+106 and 1611+343 are small, 
no higher than about 24~rad\,m$^{-2}$, which is smaller than the typical 
uncertainties in the parsec-scale RM values. We did not 
remove the effect of these small integrated RM values, since this will not
have a bearing on the interpretation of our results. On the other hand,
the integrated RMs for 2005+403 ($-171$~rad\,m$^{-2}$) and 2200+420 
($-199$~rad\,m$^{-2}$) are substantial, and we accordingly removed these RMs 
from all values in the RM maps shown for these two sources. 


\subsection{4.6--15.4~GHz, September 2007}

The data considered here were obtained as part of the same project as the
maps published by Gabuzda et al. (2014b), and were obtained on 26th September
2007. The observations and calibration procedures are described by
Gabuzda et al. (2014b). 

\subsection{5.0--15.3~GHz, September 2005} 

The data considered here were obtained on 10th September 2005 and were 
previously analyzed by Kharb et al. (2009), who describe the observations 
and calibration procedures. Their results were based on observations at 
three frequencies: 5.0, 8.4 and 15.3~GHz. 

We used final, fully calibrated UV data files kindly provided by P. Kharb
to construct Stokes $Q$, Stokes $U$, PANG, and PANGN maps.
The effect of the integrated (Galactic)
Faraday rotation was not removed, as the integrated RM (10~rad\,m$^{-2}$) 
was appreciably
smaller than the typical RM uncertainties ($\simeq 50$~rad\,m$^{-2}$).  
These essentially reproduced the
RM map of Kharb et al. (2009), but using the more accurate error estimates
given by the formula of Hovatta et al. (2012).

\subsection{8.1--15.2~GHz, June 2000}

Zavala \& Taylor (2002, 2003) present 15.2-GHz Faraday RM maps for 20 AGNs
based on VLBA observations obtained on 27th June 2000 at 8.1, 8.2, 8.4,
8.6, 12.1, 12.6 and 15.2~GHz.  We retrieved these data from the VLBA
archive and calibrated them using the same procedures as 
those described by Zavala \& Taylor (2002, 2003).  

As in Gabuzda et al (2015), we used the human eye as an initial gradient
detector; this indicated four candidates for AGNs with transverse RM 
gradients across their jets: 0212+735, 0415+379 (3C111), 1611+343 and 
2005+403.  
Our RM maps for these four sources basically reproduced the 
RM maps published by Zavala \& Taylor (2002, 2003), but applying the
more accurate error-estimation formula of Hovatta et al. (2012), and with 
the Galactic RM value removed for 2005+403.



\section{Results}

The source names,
redshifts, pc/mas values and integrated rotation measures are summarized
in Table~1.  The pc/mas were determined assuming a cosmology with $H_o = 
71$~km\,s$^{-1}$Mpc$^{-1}$, $\Omega_{\Lambda} = 0.73$ and $\Omega_{m} = 
0.27$; the redshifts and pc/mas vales were taken from the MOJAVE project 
website (http://www.physics.purdue.edu/MOJAVE/). 
Polarization maps for all of these sources at 15~GHz can be 
found in Lister \& Homan (2005) and on the MOJAVE website 
(http://www.physics.purdue.edu/MOJAVE/).

Like Motter and Gabuzda (2017), in all cases, we present total 
intensity (Stokes $I$) and RM maps made using the naturally weighted 
elliptical convolving beams as well as circular convolving beams having 
equal area; in other words, the circular beam used has a full width 
at half maximum equal to $\sqrt{(BMAJ)(BMIN)}$, where $BMAJ$ and 
$BMIN$ are the full widths at half maximum for the major and minor 
axes of the nominal elliptical restoring beam.  As is explained by Motter
and Gabuzda (2017), the maps made using 
circular convolving beams were used to test the robustness of RM structures 
visible in the maps made using the elliptical beams --- for example, an
RM gradient that seemed to be present in the original RM map but 
disappeared upon convolution with the equal-area circular beam would not 
be considered reliable (see, e.g., the case of 2155--152 presented by
Gabuzda et al. 2015). Such comparisons are especially helpful when the 
elliptical beam is very elongated. In addition, in some cases, the maps 
made using equal-area circular beams helped clarify the relationship
between structure in the RM map and the local direction of the jet.
Note that this does not bias the resulting circular beams to maximize 
the resolution in any particular direction of interest, e.g., the 
direction across the jet. 

Total intensity and RM maps made using the naturally weighted elliptical
and equivalent-area circular convolving beams are shown in 
Figs.~\ref{fig:rmmaps-BG173-1}--\ref{fig:rmmaps-ZT}, together with 
example slices in regions of visible transverse RM gradients. 
The frequency, peak and bottom contour of the intensity maps shown in
these figures are given in Table~2; the contours increase in steps of
a factor of two, and the ranges of the RM maps are 
indicated by the colour wedges shown with the maps.
The panels show the intensity maps made using the nominal elliptical
beams with the corresponding RM distributions superposed (left), the
corresponding maps made using equal-area circular beams (middle),
and slices taken along the lines drawn across the RM distributions in the
middle panels (right); the 
letter ``S'' at one end of these lines marks the side corresponding to
the starting point for the slice (a slice distance of 0). 

In all cases, we aimed to take the RM slices shown in 
Figs.~\ref{fig:rmmaps-BG173-1}--\ref{fig:rmmaps-ZT} perpendicular to the 
local jet direction.  We did not formally fit a ridge line to the jet.
When RM gradients were 
detected relatively far from the core in relatively straight jets
(0305+039, 0415+379, 2200+420), the jet direction was estimated directly
from the intensity images.  When RM gradients were detected in the region
of the VLBI core and/or inner jet (0212+735, 0300+470, 0945+408, 1502+106, 
1611+343, 2005+403), we took the slices perpendicular to the direction of 
the innermost VLBI jet visible in the 15-GHz images for the corresponding
observing epochs, taking into account the visual appearance of these maps
and the distribution of CLEAN components, as is also described by Motter
\& Gabuzda (2017). Slices in the core region were 
taken at locations near the center of the region where the gradients are 
visible, which was sometimes slightly upstream or downstream of the intensity 
peak; these locations were not particularly chosen to maximize the significance 
of these transverse RM gradients. 

RM gradients are visible across the core regions of a number of the 
sources considered here.  As is discussed by Motter \& Gabuzda (2017) 
and Wardle (2017), the polarized emission in the core region in practice 
arises in optically thin regions in the innermost jets that are blended 
with partially optically thick regions in the observed VLBI ``core.'' In
fact, as is shown by the calculations of Cobb (1993) [see also Wardle 
(2017)], the $90^{\circ}$ rotation of the polarization position angle 
associated with the optically thin--thick transition occurs near optical 
depth $\tau\simeq 6$, far upstream of the most optically thick regions in 
the observed VLBI core, located near $\tau\simeq 1$. We have accordingly
analyzed and interpreted transverse RM gradients in the core region
in the same way
as those observed farther out across the jet structures.

The statistical significances of the transverse gradients detected in our
RM maps are summarized in Table~3.  We took the significance of an RM 
gradient to be the magnitude of the difference between the RMs at the two 
ends of the gradient, divided by the uncertainty in this difference, 
taken to be the sum of the individual RM
uncertainties added in quadrature: 
\begin{eqnarray}
\textrm{Significance} & = & \frac{|RM_1 - RM_2|}{\sqrt{\sigma_{RM1}^2 + 
\sigma_{RM2}^2}}
\end{eqnarray}
Note that this approach is more
conservative that the procedure used by Hovatta et al. (2012), who compared 
the magnitude of the RM difference and the the maximum error along the 
slice; typically, our significance estimates will be about a factor 
of $\sqrt{2}$ lower, helping to ensure that we do not overestimate the
significance of the gradients. When plotting the slices in 
Figs.~\ref{fig:rmmaps-BG173-1}--\ref{fig:rmmaps-ZT} and finding the 
difference between
the RM values at two ends of a gradient, we did not include uncertainty 
in the polarization angles due to EVPA calibration uncertainty,  
since EVPA calibration uncertainty cannot introduce spurious RM gradients
(see discussions by Mahmud et al. (2009) and Hovatta et al.  (2012)).

In all cases, in both the jet and core regions, the dependence of 
$\chi_{obs}$ as a function of $\lambda^2$ is consistent with the linear 
behaviour expected for external Faraday rotation, to within the uncertainties 
in $\chi_{obs}$.

Results for each of the datasets and each of the AGNs considered here 
are summarized below.

\subsection{4.6--15.4~GHz data}

\smallskip
\noindent
{\bf 0300+470.} 
The RM maps for this source are shown in Fig.~\ref{fig:rmmaps-BG173-1} (top 
row).  The RM map made using the elliptical convolving beam (left panel) shows 
an RM gradient across the core region, whose significance is about $3\sigma$.
The RM map made using an equal-area circular beam (middle panel) likewise
shows a clear, monotonic transverse RM gradient across the core region, 
with a significance of $4.9\sigma$ (right panel).

\smallskip
\noindent
{\bf 0945+408.} 
The RM maps for this source are shown in Fig.~\ref{fig:rmmaps-BG173-1} (second 
row). Although the shift between the intensity images at the different
frequencies due to the change in the position of the VLBI core with
frequency was essentially negligible, there was a large shift between the
maps at 4.6 and 5.0~GHz and the maps at the higher frequencies, due to the 
fact that a bright knot in the inner jet was the brightest feature at 4.6 and 
5.0~GHz, while the core was the brightest feature at the higher frequencies. 
The polarization angle images at 4.6 and 5.0~GHz were shifted to correct for 
this large misalignment before the RM maps were made.

The RM map made using the elliptical convolving beam (left panel) shows
a transverse RM gradient, which has a significance of about
$3.5\sigma$. The RM map made using an equal-area 
circular beam (middle panel) shows a clear region of transverse RM gradients 
across the core, with significances of $4-5\sigma$ (right panel).

\smallskip
\noindent
{\bf 1503+106.} 
The RM maps for this source are shown in Fig.~\ref{fig:rmmaps-BG173-1} (third 
row).  The RM map made using the elliptical convolving beam (left panel)
shows a transverse RM gradient across the core region, which has a high 
significance of $7.4\sigma$, but is nevertheless somewhat uncertain due to the 
highly elliptical beam.  There is also a transverse gradient in the opposite
direction in the jet, with a significance of about $2.5\sigma$. Both of these
gradients become more clearly visible in 
the RM map made using an equal-area circular beam (middle panel), with 
significances reaching $8\sigma$ for the core region and $4-5\sigma$ for the
jet.  

\smallskip
\noindent
{\bf 2200+420.} 
The RM maps for this source are shown in Fig.~\ref{fig:rmmaps-BG173-2}.  The 
RM map made using the elliptical 4.6-GHz beam (upper row, left map) shows 
hints of a transverse RM gradient in the jet (higher RM values on the
eastern side of the jet), but it is not fully monotonic. A region with
a monotonic transverse RM gradient with a significance of about $4\sigma$
appears at the end of the detected RM distribution in the RM map made with 
the equal-area circuar beam (upper row, right map and slice), but we 
considered this gradient to be uncertain due to the small region where it 
is present at the end of the detected RM distribution.

To test the reliability of this transverse gradient, 
we made the 4.6-GHz intensity map and the RM map
using the slightly smaller 7.9-GHz elliptical beam (lower row, left map) 
and a corresponding equal-area circular beam (lower row, right map). 
This beam is about a factor of 0.6 smaller than the naturally weighted
4.6-GHz beam; use of such a beam is justified by the analysis of Coughlan 
\& Gabuzda (2016), who used Monte Carlo simulations to demonstrate that 
both Maximum-Entropy and CLEAN deconvolutions yielded reliable results for
intensity, polarization and RM images when resolving beams down to half 
the size of the full naturally weighted CLEAN beam were used.  

Transverse RM gradients in the same direction are visible across the jet 
in these slightly 
higher-resolution maps. The gradient in the region of the RM slice shown
in the top right panel of Fig.~\ref{fig:rmmaps-BG173-2} has fallen to about 
$2.3\sigma$, but a region closer to the core has RM gradients reaching 
$3-4\sigma$. Although we believe this gradient is likely real, it would
be valuable to verify this result using other multi-frequency data with 
similar resolution.

\subsection{5.0--15.3~GHz data}

\smallskip
\noindent
{\bf 0305+039 (3C78).} 
The RM maps for this source are shown in Fig.~\ref{fig:rmmaps-PK}.  The RM 
map made using the elliptical beam (left panel) shows a roughly transverse
RM gradient with a significance of about $2.7\sigma$.  The RM map made with 
an equal-area circular beam (middle panel) shows a clear, monotonic 
transverse RM gradient across this region, with a significance of 
$3.6\sigma$, whose direction is very close to orthogonal to the jet. 

\subsection{8.1--15.2~GHz data}

RM maps of these sources based on the same visibility data but with somewhat
different weighting were originally published by Zavala \& Taylor (2002, 
2003).  In all cases, our RM maps are very similar to the originally 
published maps.

\smallskip
\noindent
{\bf 0212+735.} 
The RM maps for this source are shown in Fig.~\ref{fig:rmmaps-ZT} (top row).
Both the RM maps made using the elliptical (left panel) and equal-area
circular (middle panel) beams show a very clear, monotonic transverse 
RM gradient across the core and inner jet region, with significances of 
$7-8\sigma$.

\smallskip
\noindent
{\bf 0415+379 (3C111).} 
The RM maps for this source are shown in Fig.~\ref{fig:rmmaps-ZT} (second row).
The polarization in the core and innermost jet is weak, leading to the
measurement of RM values only in the jet, well separated from the core region.
Both the RM maps made using the elliptical (left panel) and 
equal-area circular 
(middle panel) beams show a clear, monotonic transverse RM gradient across 
the jet approximately 6~mas from the core. The gradient has significances
of $3.7\sigma$ and about $6\sigma$ in the elliptical-beam and circular-beam RM
maps, respectively. 

\smallskip
\noindent
{\bf 1611+343.} 
The RM maps for this source are shown in Fig.~\ref{fig:rmmaps-ZT} (third row).
Both the RM maps made using the elliptical (left panel) and 
equal-area circular
(middle panel) beams show a clear, monotonic transverse RM gradient across
the core region. The significance of this gradient is $2.5\sigma$ in the
elliptical-beam RM map, but increases to $3.3\sigma$ in the equal-area 
circular-beam RM map. 

\smallskip
\noindent
{\bf 2005+403.} 
The RM maps for this source are shown in Fig.~\ref{fig:rmmaps-ZT} (fourth row).
Bearing in mind that the direction of the inner jet is slightly south of
east (as can be seen, for example, in the 15-GHz maps of Zavala \&
Taylor (2003) and Lister \& Homan 2005), the RM map made with the 
elliptical convolving beam shows a possible transverse gradient in the
vicnity of the core with a significance of about $5\sigma$.  The RM map 
made using an 
equal-area circular beam shows this gradient and its direction relative to 
the jet much more clearly; its
significance in the circular-beam RM map is $8-9\sigma$. The reduced clarify
of this gradient in the elliptical-beam RM map is due to the fact that, 
although the elliptical beam is not extremely elongated, its orientation is 
close to orthogonal to the jet direction. 

\begin{center}
\begin{table}
\begin{tabular}{c|c|c|c|c|}
\multicolumn{5}{c}{Table 3: Summary of transverse RM gradients}\\
\hline
Source & Beam & Location &  Width & Significance \\ 
       & Used &          & (Beams)  &  \\
0212+735 & E & Jet  & 2.0 & $8.6\sigma$\\
0212+735 & C & Jet  & 1.9 & $7.7\sigma$\\\hline
0300+470 & E& Core & 1.7 & $3.0\sigma$\\
0300+470 & C& Core & 1.8 & $4.9\sigma$\\\hline
0305+039 & E& Jet  & 0.6  & $2.7\sigma$\\
0305+039 & C& Jet  & 0.7  & $3.6\sigma$\\\hline
0415+379 & E& Jet  & 1.5  & $3.7\sigma$\\
0415+379 & C& Jet  & 1.4  & $6.1\sigma$\\\hline
0945+408 & E& Core  & 1.6 & $3.5 \sigma$\\
0945+408 & C& Core  & 1.7 & $ 4.9\sigma$\\\hline
1502+106 &E & Core  & 1.3 & $7.4\sigma$  \\
1502+106 &C & Core  & 1.4 & $8.3\sigma$ \\
1502+106 &E & Jet   & 1.0 & $2.5\sigma$ \\
1502+106 &C & Jet   & 1.0 & $4.5\sigma$\\\hline
1611+343 & E& Core & 2.0 & $ 2.5\sigma$\\
1611+343 & C& Core & 1.9 & $ 3.3\sigma$\\\hline
2005+403 & E & Jet & 1.2 & $5.0\sigma$ \\
2005+403 & C & Jet & 1.1 & $9.3\sigma$\\\hline
2200+420 & E & Jet  & 1.1 & $4.2\sigma$\\
2200+420 & C & Jet  & 1.1 & $3.5\sigma$\\ 
\hline
\end{tabular}
\end{table}
\end{center}

\section{Discussion}

\subsection{Significance of the Transverse RM Gradients}

Table~3 gives a summary of the transverse Faraday rotation measure gradients 
detected in the images presented here. The statistical significances of
these gradients typically lie in the range $3-5\sigma$, although four
of the transverse RM gradients we have detected exceed $6\sigma$.
In the case of 0305+039 and 1611+343, the transverse gradients in the RM maps 
made with their naturally weighted elliptical beams have significances 
determined using our conservative approach of $2.7\sigma$ and $2.5\sigma$, 
respectively, but these increase to $3.6\sigma$ and $3.3\sigma$, respectively, 
when the RM maps are made with a circular beam of equal area. Recall that, 
as we pointed out above, our significances will typically be about a factor 
of $\sqrt{2}$ lower than would be obtained using the approach of Hovatta 
et al. (2012) [using the maximum error rather than the two RM errors added 
in quadrature when determining the significance]; 
these two elliptical-beam significances would increase to about
$3.8\sigma$ and $3.5\sigma$, respectively, using this latter approach.
In reality, it is likely that our significances are slightly underestimated,
while those of Hovatta et al. (2012) are somewhat overestimated; the
true significances probably lie somewhere between the two, and 
we accordingly believe both of these gradients to be significant.

Thus, the Monte Carlo simulations of Hovatta et al.  (2012) and 
Murphy \& Gabuzda (2013) indicate that the probability that any of the 
transverse RM gradients in Table~3 are spurious is less than 1\%; this 
probability will 
be even lower for clear, monotonic gradients encompassing differences of 
appreciably greater than $3\sigma$ (0212+735, 0415+379, 0945+408, 1502+106,
2005+403). 

\subsection{Sign Changes in the Transverse RM Profiles}

In general, Faraday-rotation gradients can arise due to
gradients in the electron density and/or
line-of-sight magnetic field. However, since gradients in the electron density
cannot bring about changes in the sign of the Faraday rotation, a monotonic 
RM gradient encompassing an RM sign change unambiguously indicates a change 
in the direction of the line-of-sight magnetic field, such
as that due to the presence of a toroidal {\bf B} field 
component in
the region of Faraday rotation. 

Signficant sign changes are observed in the transverse RM gradients detected 
in 5 of the 9 AGNs considered here: 0300+470, 0415+479,  1502+106, 2005+403 
and 2200+420.  This supports the idea that these gradients are 
due to toroidal, possibly helical, jet {\bf B} field.
As has been noted previously (e.g. Motter \& Gabuza 2017),
the absence of a sign change in an transverse RM gradient
does not rule out the possibility that the origin of this gradient 
is toroidal field component, since RM gradients encompassing 
a single sign can sometimes be observed, depending on the
helical pitch angle and jet viewing angle.

\subsection{Core-Region Transverse RM Gradients}

As is noted in the Introduction, in the standard theoretical picture, 
the VLBI intensity ``core'' represents the intensity ``photosphere'' of 
the jet, where the optical depth is roughly unity. 
The observed VLBI core encompasses this partially optically 
thick region together with much more highly polarized optically thin 
regions in the innermost jet, with the latter dominating overall observed 
``core'' polarization.  

Various theoretical models and other high-resolution observations
support this picture. For example, Marscher et al. (2008) use the model 
of Vlahakis (2006) to explain rapid, smooth rotations of the optical 
polarization position angles as reflecting the motion of a distinct region
of polarized emission along a helical stream-line located upstream of the
observed VLBI core at millimeter wavelengths; since correlated 
polarization-angle rotations in the optical and radio have also been
observed (d'Arcangelo et al. 2009), this implies the presence of optically 
thin emission regions upstream of the 7-mm VLBI core, which Marscher
et al. (2008) suggest may actually represent the region of a recollimation 
shock, rather than a $\tau=1$ surface. Similarly, G\'omez et al. (2016)
have reported the detection of polarized emission upstream of the VLBI
core in their high-resolution observations of 2200+420.

Since the observed ``core'' polarization at centimeter and long millimeter 
wavelengths is thus actually dominated by the contributions of effectively
optically thin regions, the simplest interpretation of transverse RM gradients 
observed across a core region is that, like transverse gradients farther 
out in the jets, they reflect the presence of a toroidal {\bf B} field 
component.
Monotonic transverse RM gradients with significances of $3\sigma$ or more
are observed across the core regions of 0212+735, 0300+470, 0945+408, 1502+106, and 1611+343.   

The simulated RM maps of Broderick \& McKinney (2010) and Porth et al. (2011)
explicitly show the presence of clear, monotonic transverse RM gradients 
in core regions containing helical magnetic fields, with relativistic and 
optical depth effects only occasionally giving rise to non-monotonic behaviour 
for some azimuthal viewing angles.  In addition, the slightly non-monotonic 
behaviour displayed by some of these calculated RM profiles will be smoothed by 
convolution with a typical centimeter-wavelength VLBA beam, giving rise to
monotonic gradients of the sort reported here [see, e.g., the lower right 
panel in Fig.~8 of Broderick \& McKinney (2010)]. Therefore, when 
a smooth, monotonic, statistically significant transverse RM gradient is 
observed across the core region, it is justified to interpret this 
as evidence for helical/toroidal jet {\bf B} fields on the corresponding
scales; this is particularly so given that the $90^{\circ}$ rotation in the
polarization angle associated with the optically thin/thick transition does
not occur until optical depths $\tau\simeq 6$ (Cobb 1993, Wardle 2017).
 
\subsection{RM-Gradient Reversals}

We have detected evidence for distinct regions with transverse Faraday 
rotation
gradients oriented in opposite directions in 1502+106. Similar reversals 
in the directions of the RM gradients in the core region and inner jet 
have been reported for 0716+714, 0923+392, 1749+701 2037+511 (Mahmud et 
al. 2013, Gabuzda et al. 2014b). 

Our results for 2200+420 are also of interest here. Motter \& Gabuzda (2017)
presented a 1.4--1.7-GHz VLBA RM map of this same object for epoch August 2010,
about three years after the 4.6--15.4~GHz observations considered here. Their
map also shows an RM gradient across the core of their image, but in the
opposite direction to the one presented here. Further, G\'omez et al.
(2016) have produced a high-resolution RM map based on joint analysis of 
22-GHz RadioAstron space-VLBI data and ground-based VLBI data obtained at 
15 and 43~GHz in November 2013, which likewise shows a transverse RM gradient 
in the vicinity of the high-frequency VLBI core, in the same direction as 
that observed by Motter et al. (2016). This raises the possibility that the
predominant direction of the transverse RM gradients in 2200+420 may vary 
with time, as has also been observed for 1803+784 (Mahmud et al. 2009) and
0836+710 (Gabuzda et al. 2014).

As is discussed by Mahmud et al. (2009, 2013), one reasonable interpretation
of both of these phenomena (RM-gradient reversals along the jet and in time)
is a picture with a nested helical field structure,
with opposite directions for the azimuthal field components 
in the inner and outer regions of helical field. The total observed
Faraday rotation in the vicinity of the AGN jet includes contributions
from both these regions, and a change in the direction 
of the net observed RM gradient could be due to a change in dominance from 
the inner to the outer region of helical field, in terms of their
overall contribution to the observed Faraday rotation. One possible
physical picture giving rise to such a nested helical-field structure
is described by Christodoulou et al. (2016).
Lico et al. (2017) have used this type of model to explain changes in the
sign of the 15--43~GHz VLBA core RM of Mrk~421 in a similar way.

\subsection{Transverse RM Gradients and AGNs with ``Spine--Sheath''
Magnetic-Field Structures} 

The objects observed in September 2007 at 4.6--15.4~GHz (0300+470,
0945+408, 1502+106 and 2200+420) were part of the same sample of sources 
displaying ``spine--sheath''polarization structures considered by
Gabuzda et al. (2014b). 
One possible origin for this polarization structure is a helical
jet {\bf B} field, with the sky projection of the helical field 
predominantly orthogonal to the jet near the jet axis and 
predominantly longitudinal near the jet edges. This suggests that
transverse RM gradients may also be common across these jets.

In all, 22 AGNs with spine--sheath polarization structure were observed
as part of this experiment (BG173): 12 sources on 26th September 2007 and 10 
sources on 27th September 2007.  This is to our knowledge the only set of 
VLBI observations aimed at Faraday rotation studies in which the target AGNs 
were selected based on the hypothesis that they were good candidates for 
AGNs with RM distributions showing transverse RM gradients (with jets 
carrying helical magnetic fields).

Taking the results presented here together with those of Gabuzda et al. 
(2014b), 12 of these 22 AGNs (i.e., about 55\%)
displayed statistically significant transverse RM gradients across their jets, 
while the remaining 10 AGNs did not show significant transverse RM gradients.  
In contrast, the re-analysis of the multi-frequency data of Zavala \& Taylor 
(2002, 2003, 2004) carried out by Gabuzda et al. (2015) and in the current
paper indicates the presence of statistically significant 
transverse RM gradients in 6 out of 40 AGNs (i.e., about 15\%).  
Thus, we have found a considerably higher fraction of AGNs displaying partial 
or full ``spine--sheath'' transverse polarization structures to display
firm evidence for transverse RM gradients, compared to the AGN sample of
Taylor (2000), which was selected to have 15-GHz flux greater than 2~Jy and
declinations greater than $-10^{\circ}$. 
This supports the hypothesis that 
both the ``spine--sheath'' polarization structure and the relatively high
incidence of transverse RM gradients among the 22 AGNs considered by 
Gabuzda et al. (2014b) and in the current paper is due to the fact that
these AGN jets carry helical {\bf B} fields. 
Another possibility is that our criterion that the sources display 
``spine--sheath'' polarization structure has essentially selected a set 
of sources with transversely resolved linear polarization structure, whose 
transverse RM structures are likewise relatively well resolved, making it 
easier to detect transverse RM gradients in these sources; in this case,
this would suggest that a relatively large fraction of {\em all} 
AGN might display transverse RM gradients.  The high fraction of  
transverse RM gradients observed in the ``spine--sheath'' sources that 
display sign changes in the RM (9 of 12) also supports the idea that these 
gradients are due to toroidal or helical jet {\bf B} fields, since an RM 
sign change can only be explained by a change in the direction of the 
line-of-sight {\bf B} field, not a change in the electron density.

\section{Conclusion}

We have presented new polarization and Faraday RM measurements of 4
AGNs based on 4.6--15.4~GHz observations with the VLBA, 
together with a reanalysis of RM maps for 4 AGNs published
previously by Zavala \& Taylor (2002, 2003) and an RM map for the radio
galaxy 0305+039 (3C78) published previously by Kharb et al. (2009). 
All 9 of these AGNs display Faraday rotation measure (RM) 
gradients across their core regions and/or jets with conservatively
estimated statistical significances of at least $3\sigma$. 

One of the AGNs considered here  --- 1502+106 ---
shows evidence for distinct regions with transverse Faraday rotation
gradients oriented in opposite directions, and another --- 2200+420 --- for
changes in the direction of its transverse RM gradients with time. 
Similar reversals in the directions of the RM gradients in the core
regions and inner jets have been observed for 4 other AGNs (Mahmud et al.
2013, Gabuzda et al. 2014), while reversals in the directions of the RM 
gradients with time have been observed for 2 other AGNs (Mahmud et al.
2009, Gabuzda et al. 2014b). These have been 
interpreted as evidence for a nested helical field structure, with 
the inner and outer regions of helical field having oppositely directed  
azimuthal components. 

Our results for 0300+470, 0945+408 and 1503+106 support the conclusion
of Gabuzda et al. (2014b) that statistically signficant transverse 
RM gradients
are common across the VLBI jets of AGNs displaying transverse
polarization structures with a ``spine'' of transverse magnetic field
and a ``sheath'' of longitudinal magnetic field. 
This is natural if both the transverse polarization
structure and the transverse RM gradients in these AGNs have their origin
in a helical {\bf B} field located in the jets and in their immediate
vicinity, as would come about through the winding up of an initial 
longitudinal field component by the rotation of the central black hole and
its accretion disk.

\section{Acknowledgements}

Partial funding for this research was provided by the Irish Research 
Council (IRC). We thank A. Reichstein and S. Knott for their work on
preliminary calibration and preliminary imaging of some of the data
considered here.

\end{document}